\documentclass[preprint,showpacs,amsmath,amssymb,letterpaper]{revtex4}

\usepackage{graphicx}% Include figure files
\usepackage{dcolumn}% Align table columns on decimal point
\usepackage{bm}% bold math

\begin{document}

%\preprint{APS/123-QED}

\title{Comparisons and Combinations of Reactor and Long-Baseline
  Neutrino Oscillation Measurements}

\author{K. B. M. Mahn}
\email{kendallm@phys.columbia.edu}
\author{M. H. Shaevitz}
\email{shaevitz@nevis.columbia.edu}
\affiliation{Department of Physics, Columbia University, New York, NY 10027}
\date{\today}

\begin{abstract}
We investigate how the data from various future
neutrino oscillation
experiments will constrain the physics parameters for a three active
neutrino mixing model.  The investigations properly account for the
degeneracies
and ambiguities associated with the phenomenology as well as estimates of
experimental measurement errors.  Combinations of various reactor measurements 
with 
the expected J-PARC (T2K) and NuMI offaxis (Nova) data, both with and without 
the increased flux associated with proton driver upgrades, 
are considered. The studies show how combinations of reactor and offaxis 
data can resolve degeneracies (e.g. the $\theta_{23}$ degeneracy)
 and give more precise information
on the oscillation parameters.  A primary purpose of this investigation
is to establish the parameter space regions where CP violation can be 
discovered and where the mass hierarchy can be determined. 
We find that, even with augmented flux from proton drivers, such measurements
demand that $\sin^2 2\theta_{13}$ be fairly large and in the range where 
it is measurable by reactor experiments. 

\end{abstract}

% PACS, the Physics and Astronomy
                             % Classification Scheme.
%\keywords{Suggested keywords}%Use showkeys class option if keyword
                              %display desired
\maketitle

The worldwide program to understand neutrino oscillations and determine the
mixing parameters, CP violating effects, and mass hierarchy will require a
broad combination of measurements. Progress in the past associated with
solving the solar and atmospheric neutrino puzzles took a full suite of
experiments to isolate and understand the phenomenology. Each additional measurement helped define the direction of future studies. One would
expect a similar chain for the current goals where the program grows as
information is obtained. This study attempts to see how various present
proposals for next generation experiments (including two detector reactor and
accelerator based long-baseline experiments) compare to and complement each
other. A particular emphasis is on combining experiments to give improved
physics parameter determination. As in the past, the best constraints on the
phenomenology come from combining data from different processes and setups.

\section{Procedure for this study}

For a three active neutrino scenario, neutrino oscillations are described by
six physics parameters: $\theta_{13},\theta_{12},\theta_{23},\Delta m_{12}%
^{2},\Delta m_{23}^{2},$ and the CP violation phase, $\delta$. In addition, a
full description requires knowing the hierarchy of mass state 3 relative
to 1 and 2, \textit{i.e. }the sign of $\Delta m_{23}^{2}$. 
(See Ref.\cite{Huber:2004ug,Barger} for a description of
three neutrino oscillation phenomenology and current results of global fits.)

The oscillation probability up to second order for reactor and
long-baseline measurements is given by \cite{Huber:2004ug}

\begin{equation}
P_{reactor}  \simeq \quad \sin^2 2 \theta_{13} \,
\sin^2 \Delta + \alpha^2 \, \Delta^2 \, \cos^4 \theta_{13}
\, \sin^2 2 \theta_{12} , \label{OscProb}
\end{equation}
%

%\begin{equation}
%P_{reactor}  =1-P(\nu_{e}\rightarrow\nu_{e})
%\simeq\sin^{2}2\theta_{13}
%\sin^{2}\left(  1.27\Delta m_{31}^{2}L/E\right) +\cdots \label{OscProb}
%\end{equation}

%\begin{align}
%P_{long-baseline}=P(\nu_{\mu}\rightarrow\nu_{e}) 
%\simeq \sin^{2}2\theta_{13}\sin^{2}\theta_{23}
%\sin^{2}\left(  1.27\Delta m_{31}^{2}L/E\right)   \nonumber \\
%   +\text{ Other terms with CP viol. and matter effects}
%\end{align}

%
\begin{eqnarray}
P_{long-baseline} & \simeq & \sin^2 2\theta_{13} \, \sin^2 \theta_{23}
\sin^2 {\Delta} \nonumber \\
& \mp &  \alpha\; \sin 2\theta_{13} \, \sin\delta_{CP}  \, \cos\theta_{13} \sin
2\theta_{12} \sin 2\theta_{23}
\sin^3{\Delta} \nonumber \\
&+&  \alpha\; \sin 2\theta_{13}  \, \cos\delta_{CP} \, \cos\theta_{13} \sin
2\theta_{12} \sin 2\theta_{23}
 \cos {\Delta} \sin^2 {\Delta} \nonumber  \\
&+& \alpha^2 \, \cos^2 \theta_{23} \sin^2 2\theta_{12} \sin^2 {\Delta} 
\label{equ:beam}
\end{eqnarray} 
with $\alpha \equiv \Delta m_{21}^2 / \Delta m_{23}^2$ and $\Delta \equiv \Delta m^2_{31}L / (4E_\nu)$.

 In the investigations shown here, the full formulae for the
oscillation probability have been used as incorporated in a computer program
developed by S. Parke \cite{parke}. The higher order corrections for the 
reactor
probability are quite small for the distances on the scale of the reactor experiments considered here, and so,
approximately, a measurement of
$P_{reactor}$ directly constrains the mixing parameter
$\theta_{13}$. On the other hand, the full expression for the long-baseline
probability introduces many degeneracies and correlations between the physics
parameters $\theta_{23}$ and $\delta_{CP}$, plus the mass hierarchy through
matter effects even before experimental uncertainties are taken into 
account \cite{Huber:2004ug,Minakata}.
Therefore, a measurement of $P_{long-baseline}$ corresponds to sizable 
regions in the physics parameter space.
 
Of the six
parameters, we assumed for this study that $\theta_{12},\theta_{23},\Delta
m_{12}^{2},$ and $\Delta m_{23}^{2}$ are known to the precision expected from
either the current program (Super-K, Minos and CNGS) or the future program
(Nova and T2K), as shown in Table \ref{inputs}. This leaves for
determination $\theta_{13},\delta$, and the
mass hierarchy, which are the subject of this study. 

The technique we will use for this study is as follows. We generate
data according to the experimental expectations with the underlying
physics assumptions and assumed parameters. We then explore the 
parameter space for one of the measurement parameters in terms of
$\Delta \chi^{2}$ values.  For a given set of parameters, a $\chi^{2}$ 
value is found by comparing the prediction with these parameters to 
the orginally generated data.  The difference between this $\chi^{2}$ 
value and the one calculated with the original parameters is the 
$\Delta \chi^{2}$ value.  In each case, we will explore the change due 
to the variation of only one
parameter. However, estimating a correct sensitivity depends on the 
choices of all six of the
parameters described above. We handle these inherent ambiguities by
choosing the most conservative (lowest) $\Delta \chi^{2}$ value for 
the parameter being tested by marginalizing over all the other
parameters. For example,
if we are studying $\theta_{13}$, for each value of $\theta_{13}$
studied, the other parameters $\delta_{CP},\Delta m_{23}^{2}$
(including sign) and $\theta_{23}$ are allowed to vary within the
allowed ranges. This sometimes results in discontinuities, where one
hierarchy becomes worse than the other in a certain region.
The best choices are determined by minimizing the $\Delta \chi^{2}$ 
value using the Minuit\cite{minuit} program.

The other experimental inputs for the study are given in Table \ref{ExpSense}
and are derived from estimates of the measurement sensitivities. For these
sensitivities, we have taken the values from the given experiment's
estimates without further study. Three types of two detector reactor
experiments are considered corresponding to a small (Double-CHOOZ, 
\cite{Ardellier:2004ui}), medium (Braidwood \cite{braidwood}, Daya Bay \cite{dayabay}), or 
large (MiniBooNE size) detector reactor $\overline{\nu}_{e}$ measurements. 
The sensitivities for the reactor experiments are scaled from the $\sin^{2}%
2\theta_{13}$ 90\% C.L. limits at $\Delta m_{23}^{2}=2.5\times10^{-3}$
eV$^{2}$ for a null oscillation scenario. 
In terms of integrated luminosity defined as detector mass [tons] $\times $
thermal reactor power [GW]$\times $ running time [years], the three options
correspond to 300, 3000, and 20,000 ton$\cdot $GW$\cdot $yrs  for the small,
medium, and large reactor scenarios. The assumed 90\% C.L. limits on $\sin
^{2}2\theta _{13}$ for the three scenarios are then given by 0.03, 0.01, and
0.005, for small, medium, and large detector reactor experiments, respectively.

\begin{table*}[ptb]
%\begin{ruledtabular}
\begin{tabular}
[c]{|cccc|}
\hline
Parameter & \multicolumn{1}{c|}{Value} & \multicolumn{1}{c|}{Current Uncertainty (1$\sigma$) }
& Future Uncertainty (1 $\sigma$) \\\hline
$\sin^{2}2\theta_{23}$ & \multicolumn{1}{c|}{1.0} & \multicolumn{1}{c|}{0.1
(Super-K)} & 0.01 (T2K)\\
$\Delta m_{23}^{2}$(eV$^{2}$) & \multicolumn{1}{c|}{$2.5\times10^{-3}$} &
\multicolumn{1}{c|}{$0.55\times10^{-3}$ (Super-K)} & $0.1\times10^{-3}$
(T2K)\\
$\theta_{12}(\deg)$ & \multicolumn{1}{c|}{32.31} & \multicolumn{1}{c|}{2.55 (global solar fit)} & --\\
$\Delta m_{12}^{2}$(eV$^{2}$) & \multicolumn{1}{c|}{$7.9\times10^{-5}$} &
\multicolumn{1}{c|}{$0.55\times10^{-5}$ (global solar fit)} & --\\
\hline
\end{tabular}
\caption{Current and future uncertainty estimates on oscillation
parameters from Super-Kamiokande \cite{Ashie:2004mr}, T2K \cite{t2kloi}, and
a global solar fit by Kamland \cite{Araki:2004mb}. The studies here use the 
above values
except the central value of $\sin^{2}2\theta_{23}$ is assumed to be 0.95 (see text).}
\label{inputs}
%\end{ruledtabular}
\end{table*}

Two offaxis 
long-baseline experiments are considered, J-PARC to Super-K\cite{t2kloi} (T2K) 
and the NuMI offaxis proposal\cite{Nova} (Nova). For these experiments,
the uncertainties are scaled from the expected number of events given in the Nova 
Proposal for 10km offaxis \cite{novaproposal} and a talk by Nakaya given at NOON 2003 \cite{Nakaya:2003qh}. The given
uncertainties include statistical errors associated with the background and
signal for a 5 year data run but do not include any systematic uncertainty. 
In the studies presented here, results are given for various data running 
periods
with the nominal beam rates corresponding to Table \ref{ExpSense}. 
In addition, some results are given for upgraded beams with five times the flux
as would be appropriate for the T2K Phase II experiment or the Nova experiment
with a new proton driver. We consider the different cases of running only with neutrinos, or with neutrino and antineutrino running with the number of years of neutrino and antineutrino running specified for each plot. The uncertainty on the $\theta_{23}$ 
parameter can have a significant effect
on the long-baseline measurements since the quantity that is constrained as
given in Table \ref{inputs} is $\sin^{2}2\theta_{23}$, and the parameter that
modulates the long-baseline oscillation probability is $\sin^{2}\theta_{23}$.
This can lead to a 60\%
 uncertainty in the oscillation probability for $\sin^2 2\theta_{23} = 0.95$.

\begin{table*}[ptb]
\begin{center}
\noindent%
\begin{tabular}
[c]{|cclcc|}\hline
%& \multicolumn{1}{c|}{\qquad\quad} 
%& \multicolumn{1}{c|}{\qquad} 
\multicolumn{1}{|l}{} 
& \multicolumn{1}{|l|}{Basis of} &
\multicolumn{3}{c|}{Osc. Prob. and $\sigma$ for $\sin^{2}2\theta_{13}=$}
\\\cline{3-5}%
\multicolumn{1}{|l}{Experiment} & 
\multicolumn{1}{|l|}{Estimate} & \multicolumn{1}{c|}{0.02} &
\multicolumn{1}{c|}{0.05} & \multicolumn{1}{c|}{0.10} \\\hline

\multicolumn{1}{|l}{Reactor ($E_{\nu}=3.6$ MeV)} &  
\multicolumn{1}{|l|}{$\sin^{2}2\theta_{13}^{Limit}$} & 
\multicolumn{1}{c|}{} & \multicolumn{1}{c|}{} & \\ 

\multicolumn{1}{|c}{Small 1.05 km } & 
%\multicolumn{1}{c|}{1.05 km} &
\multicolumn{1}{|l|}{$0.03@90\%CL$} & \multicolumn{1}{c|}{$0.013\pm0.012$} &
\multicolumn{1}{c|}{$0.033\pm0.012$} & \multicolumn{1}{c|}{$0.064\pm0.012$} \\
\multicolumn{1}{|c}{Medium 1.8 km } & 
%\multicolumn{1}{c|}{1.8 km} &
\multicolumn{1}{|l|}{$0.01@90\%CL$} & \multicolumn{1}{c|}{$0.022\pm0.006$} &
\multicolumn{1}{c|}{$0.052\pm0.006$} & $0.102\pm0.006$\\
\multicolumn{1}{|c}{Large 1.8 km } & 
%\multicolumn{1}{r|}{1.8 km} & right aligned + line
%\multicolumn{1}{c|}{1.8 km} &
\multicolumn{1}{|l|}{$0.005@90\%CL$} & \multicolumn{1}{c|}{$0.022\pm0.003$} &
\multicolumn{1}{c|}{$0.052\pm0.003$} & $0.102\pm0.003$\\\hline

\multicolumn{1}{|l}{T2K ($E_{\nu} = 650$ MeV)}
%& \multicolumn{1}{|l|}{N$_{events}^{5yrs}$:} 
& \multicolumn{1}{|l|}{Number of events in 5 yrs:} 
& \multicolumn{1}{c|}{}
& \multicolumn{1}{c|}{} & \\

$\left\langle L\right\rangle =295$ km & 
\multicolumn{1}{|l|}{} &
 \multicolumn{1}{c|}{} & \multicolumn{1}{c|}{} & \\

$\nu$ & \multicolumn{1}{|l|}{105.0 signal /
  17.8 bkgnd} &
\multicolumn{1}{c|}{$0.010\pm0.003$} & \multicolumn{1}{c|}{$0.023\pm0.004$} &
$0.044\pm0.006$\\
$\overline{\nu}$ & \multicolumn{1}{|l|}{30.8 signal /
10.2 bkgnd} & \multicolumn{1}{c|}{$0.009\pm0.007$} &
\multicolumn{1}{c|}{$0.020\pm0.008$} & \multicolumn{1}{c|}{$0.038\pm0.010$} 
\\\hline

\multicolumn{1}{|l}{Nova ($E_{\nu}=2.1$ GeV)} &
%\multicolumn{1}{|l|}{N$_{events}^{5yrs}$:} 
\multicolumn{1}{|l|}{Number of events in 5 yrs:} 
& \multicolumn{1}{c|}{} & \multicolumn{1}{c|}{} & \\

$\left\langle L\right\rangle =810$ km & 
\multicolumn{1}{|l|}{} &
\multicolumn{1}{c|}{} & \multicolumn{1}{c|}{} & \\
$\nu$ & \multicolumn{1}{|l|}{227.4 signal / 39.0 bkgnd}
& \multicolumn{1}{c|}{$0.010\pm0.002$} & \multicolumn{1}{c|}{$0.024\pm0.003$}
& $0.045\pm0.004$\\
$\overline{\nu}$ & \multicolumn{1}{|l|}{109.0 signal / 18.5
bkgnd} & \multicolumn{1}{c|}{$0.008\pm0.003$} & \multicolumn{1}{c|}{$0.017\pm
0.005$} & \multicolumn{1}{c|}{$0.032\pm0.006$} \\\hline
\end{tabular}
\end{center}
\caption{Estimates of the experimental uncertainties for
future oscillation experiments. The last three columns indicate the oscillation probability and respective error for three different values of $\sin^{2}2\theta_{13}$. For the long-baseline experiments, the number of events and sensitivity is given for 5 years of neutrino mode and 5 years of antineutrino mode separately. The given
uncertainties include statistical errors associated with the background and
signal for a 5 year data run, but do not include systematic
uncertainty.  $\Delta m_{23}^{2}=2.5\times10^{-3}$eV$^{2}$ for all
estimates and additionally, $\sin^{2}2\theta_{13}=0.1,$
$\delta_{CP}=0$ for the long baseline experiment's event rates. }%
\label{ExpSense}%
\end{table*}

For the studies presented in this report, 
the uncertainties due to the variations of
$\theta_{23}$, $\Delta m_{23}^{2}$, and the mass hierarchy are
included. In the
$\overline{\nu}$ data, there is a 5\% to 10\% contamination of $\nu$
induced events in the sample.  For the studies presented here, 
we use 5\% for Nova, and
10\% for T2K, as estimated by the experiment.  To summarize,  
the following list gives the parameters used for the various
fit results and the values and typical uncertainties for the parameters
that are varied:

\begin{itemize}
\item $\Delta m_{23}^{2}$ $($ $\approx\Delta m_{13}^{2})=2.5\times 10^{-3}$ 
eV$^{2}
$ and allowed to vary within its future expected uncertainty 
(typical $\sigma =0.1\times
10^{-3}$ eV$^{2}$). We include the hierarchy ambiguity associated with 
$\Delta m_{23}^{2}$ being  $\,<0$ or $>0$.

\item $\sin ^{2}2\theta _{23}=0.95$ and allowed to vary within its
future expected
uncertainty (typical $\sigma =0.01$). We include the ambiguity 
associated with
$\theta _{23}$ being $\,<45^{\circ }$ or $>45^{\circ }$.

\item $\Delta m_{12}^{2}$ held fixed at $7.9\times 10^{-5}$ eV$^{2}$.

\item $\theta _{12}$ held fixed at $32.31^{\circ }$.

\item $\delta _{CP}$ allowed to vary between $0^{\circ }$ to $360^{\circ }$.

\item $\sin ^{2}2\theta _{13}$ allowed to vary between $0.0$ and $0.3$.
\end{itemize}

\section{90\% C.L. allowed range of $\theta_{13}$}

The $\theta_{13}$ mixing angle is as yet undetermined in the current
neutrino oscillation phenomenology. It is known to be smaller than the
other mixing angles. The size of $\theta_{13}$ is an important ingredient in
constraining models of neutrino masses and mixing, such as attempts to relate
the quark and lepton mixings. The size of this mixing angle also has important
implications for long-baseline $\nu_{e}$ appearance measurements because it
scales the size of the oscillation probability.

\begin{figure}[ptb]
\begin{center}
\scalebox{.9}{\includegraphics[
width=7.0in
]{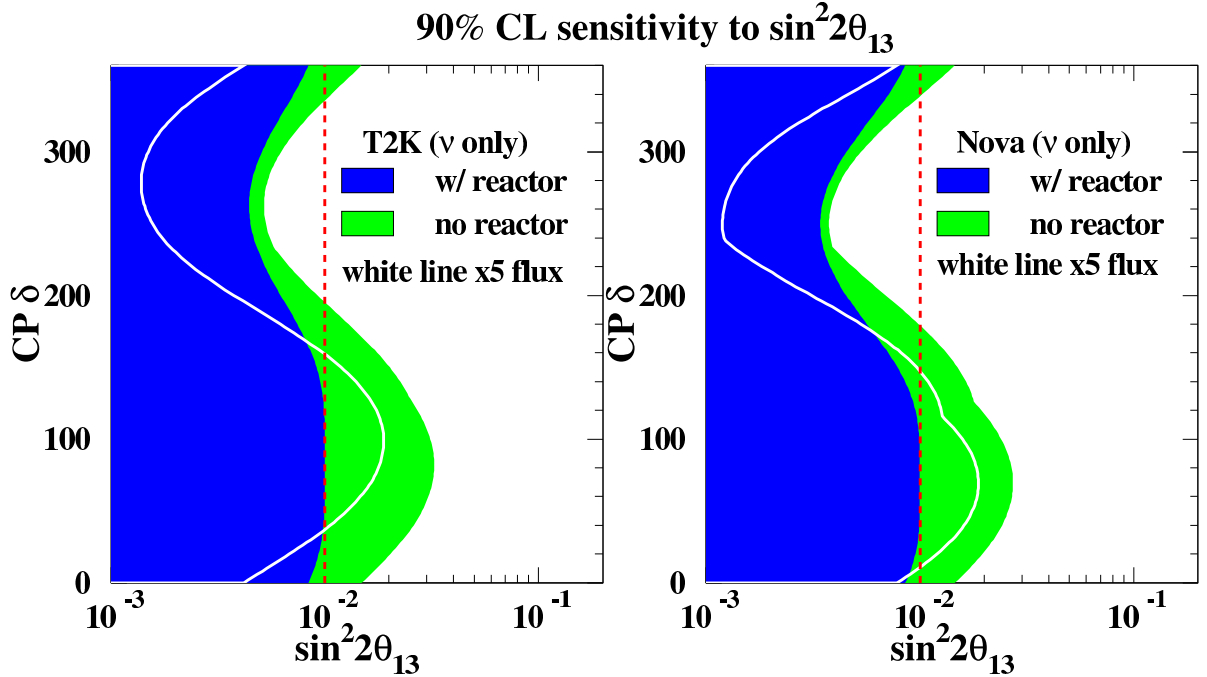}}
\end{center}
\caption{90\% C.L. upper limit regions for various oscillation
measurements for an underlying null oscillation scenario where $\sin^{2}%
2\theta_{13}=0$ ($\sin^{2}2\theta_{23}=0.95\pm0.01$, 
$\Delta m^{2}=2.5 \pm 0.1 \times10^{-3}$ eV$^{2}$ and $\delta_{CP}=0^{\circ}$). 
The left 
(right) plot is for the T2K (Nova)
long-baseline experiment. The grey region is the 90\%
C.L. allowed region for the long-baseline experiments for a three year
neutrino only run with nominal beam rate. The white line is the limit of the 90\% C.L. allowed region for a three year, neutrino only run at $5\times$ the nominal beam rate. The black region gives
the combination of three year long-baseline runs with a medium reactor
measurement. The vertical dashed line indicates the 90\% CL upper limit for a medium reactor experiment alone.}%
\label{cpvs13null}% 
\end{figure}

As an indication of how well a given measurement can constrain the value of
$\theta_{13}$, Figure \ref{cpvs13null} (left: T2K, right: Nova) shows the 90\%
C.L. allowed regions associated with measurements of a null oscillation
scenario where the true value of $\sin^{2}2\theta_{13}$ is equal to zero. 
 The grey region
(white curve) is the 90\% C.L. allowed region for the two long-baseline
experiments for a three year neutrino only run with the nominal ($\times5$)
beam rate. Combining the long-baseline and medium reactor measurement gives
the improved black region. 

\begin{figure}[ptb]
\begin{center}
\includegraphics[
width=6.5in
]{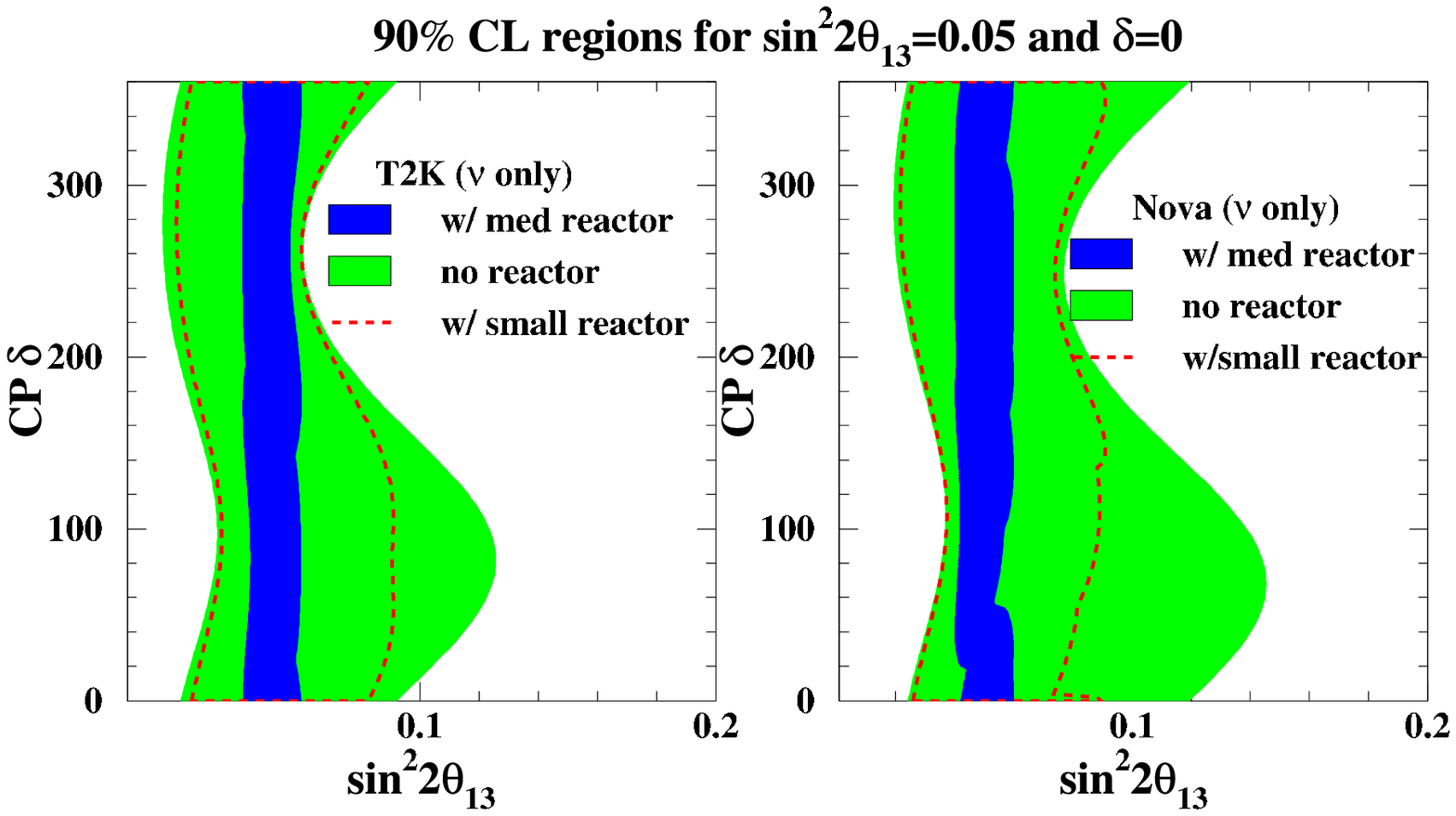}
\end{center}
\caption{90\% C.L. regions for underlying oscillation parameters of $\sin
^{2}2\theta_{13}=0.05$, $\sin^{2}2\theta_{23}=0.95\pm0.01$, 
$\Delta m^{2}=2.5\pm 0.1\times10^{-3}$ eV$^{2}$ and $\delta_{CP}=0^{\circ}$.
 The grey regions are for the T2K (left plot) or Nova (right
plot) experiments for three years of neutrino running. The black regions are 
the
90\% C.L. allowed regions for a combined medium reactor plus long-baseline
analysis. The dashed lines indicate how the combined measurement would
degrade with the small reactor sensitivity.}%
\label{dcp2}%
\end{figure}

If $\theta_{13}$ is large enough, then positive signals will be observed by
the experiments. Under these circumstances, the goal would be to make the best
determination of the mixing parameter. Figure \ref{dcp2} shows the 90\%
C.L. regions that would be obtained for an underlying scenario where $\sin
^{2}2\theta_{13}=0.05$. As shown, a long-baseline only measurement will not
determine the mixing parameter $\theta_{13}$ very well with an allowed region
that spans from 0.025 to over 0.11. On the other hand, a reactor experiment
with at least the medium scale sensitivity measures $\sin^{2}2\theta_{13}$ to
about 10\% or $\theta_{13}$ to $\pm0.4^{\circ}$.

As seen from the figures, the reactor measurements are very efficient at
constraining the value of $\theta_{13}$. Even a small reactor experiment
can probe for an early indication if the value is sizable. The large
reactor experiment has sensitivity comparable to planned long-baseline
experiments and the medium scale experiment can measure values in the range
for $\sin^{2}2\theta_{13}>0.02$ at the 10\% to 30\% level. 
As will be seen in
later plots, studies of CP violation and matter effects over the next decade
are only possible if $\sin^{2}2\theta_{13}$ is significantly larger than about
0.01. A small or medium scale reactor experiment can establish if these
studies will be possible and, if they are, add additional information for determining the mixing parameters.

\section{Resolution of the $\theta_{23}$ Degeneracy}

The mixing angle $\theta_{23}$ is an important parameter in developing an
understanding of the mixing matrix and for proceeding with a determination of
$\theta_{13}$. In many theoretical models, $\theta_{23}$ is not expected to be
45$^{\circ}$ and the difference from this value, both in sign and magnitude, 
may lead to a deeper understanding of the mixing. 

Information on the value of $\theta_{23}$ has been obtained from $\nu_{\mu}$
disappearance measurements in the atmospheric $\Delta m^{2}$ region, such as 
the Super-K and K2K experiments. These experiments restrict the
allowed region of $\sin^{2}2\theta_{23}$. Unfortunately, a single value of 
$\sin^{2}2\theta_{23}=a$ corresponds to two possible solutions for
$\theta_{23}$, $\frac{1}{2}\sin^{-1}\left(  \sqrt{a}\right)  $ or $\frac{\pi
}{2}-\frac{1}{2}\sin^{-1}\left(  \sqrt{a}\right)  $. The current Super-K
measurement of $\sin^{2}2\theta_{23}=1.00\pm0.1$ corresponds to values of
$\theta_{23}=45^{\circ}\pm9.22^{\circ}$. For the determination of $\theta_{13}$ 
using a
long-baseline $\nu_{\mu}\rightarrow\nu_{e}$ appearance measurement, this
ambiguity presents a problem since the oscillation probability is proportional
to $\sin^{2}\theta_{23}$, as shown in Equation \ref{OscProb}. The present
Super-K measurement would correspond to a change in the T2K or Nova
oscillation probability of about 63\%, for a change in $\theta_{23}$ 
from 35.78$^{\circ}$ to 54.22$^{\circ}$.

\begin{figure}[ptb]
\begin{center}
\includegraphics[
width=6.5in
]{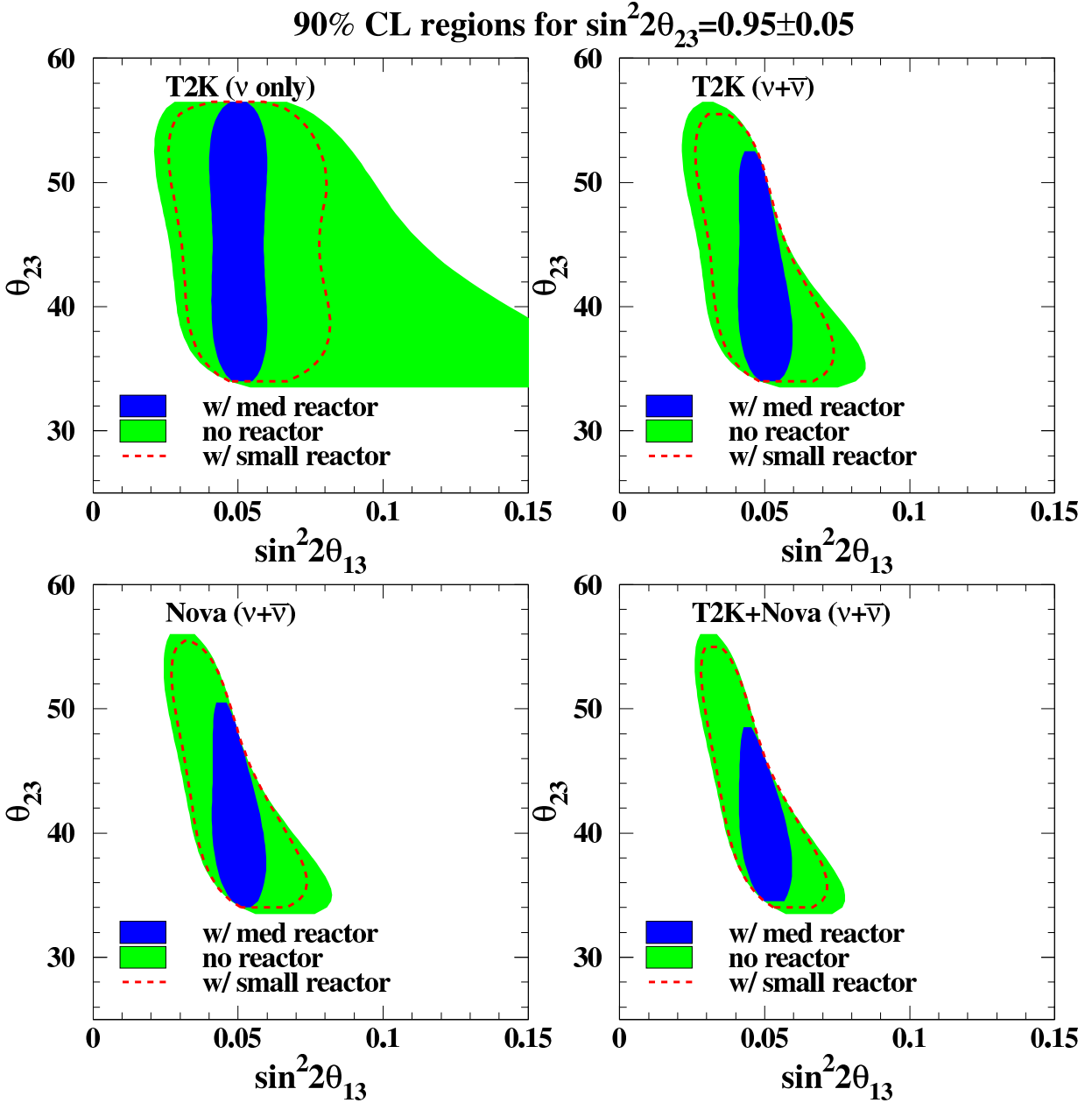}
\end{center}
\caption{90\% C.L. allowed regions for simulated data with underlying
oscillation parameters of $\sin^{2}2\theta_{13}=0.05$, 
$\theta_{23}=38.54^{\circ}$,
$\Delta m^{2}=2.5\pm 0.1\times10^{-3}$ eV$^{2}$ and $\delta_{CP}=270^{\circ}$. 
The analysis
includes the restriction that $\sin^{2}2\theta_{23}=0.95\pm0.05$. The grey
regions are for various long-baseline combinations of the T2K and Nova
experiments for three year running periods. The black regions are the 90\%
C.L. allowed regions for a combined medium reactor plus long-baseline
analysis. The dashed lines indicate a combined analysis with a small
reactor measurement.}%
\label{23vs13_060}%
\end{figure}

This ambiguity in the determination of $\theta_{23}$ is difficult for a
long-baseline $\nu_{\mu}\rightarrow\nu_{e}$ appearance measurement to resolve,
but can be well addressed with a combination of reactor and long-baseline 
measurements (see Ref. \cite{Minakata2002,Minakata:2004pg}).
Figure \ref{23vs13_060} shows examples of different combinations of
long-baseline results with (black regions) or without (grey regions) the
inclusion of a medium scale reactor measurement. (The dashed curve is for
the inclusion of a small, Double-CHOOZ type measurement.) For this analysis,
$\sin^{2}2\theta_{23}=0.95\pm0.05$, $\sin^{2}2\theta_{13}=0.05$, and
$\delta_{CP}=270^{\circ}.$ 
As shown, the medium scale reactor data resolves this degeneracy to some degree
when 
combined with antineutrino running, however a small (Double-CHOOZ) reactor 
experiment does not at all.

\begin{figure}[ptb]
\begin{center}
\includegraphics[
width=6.5in
]{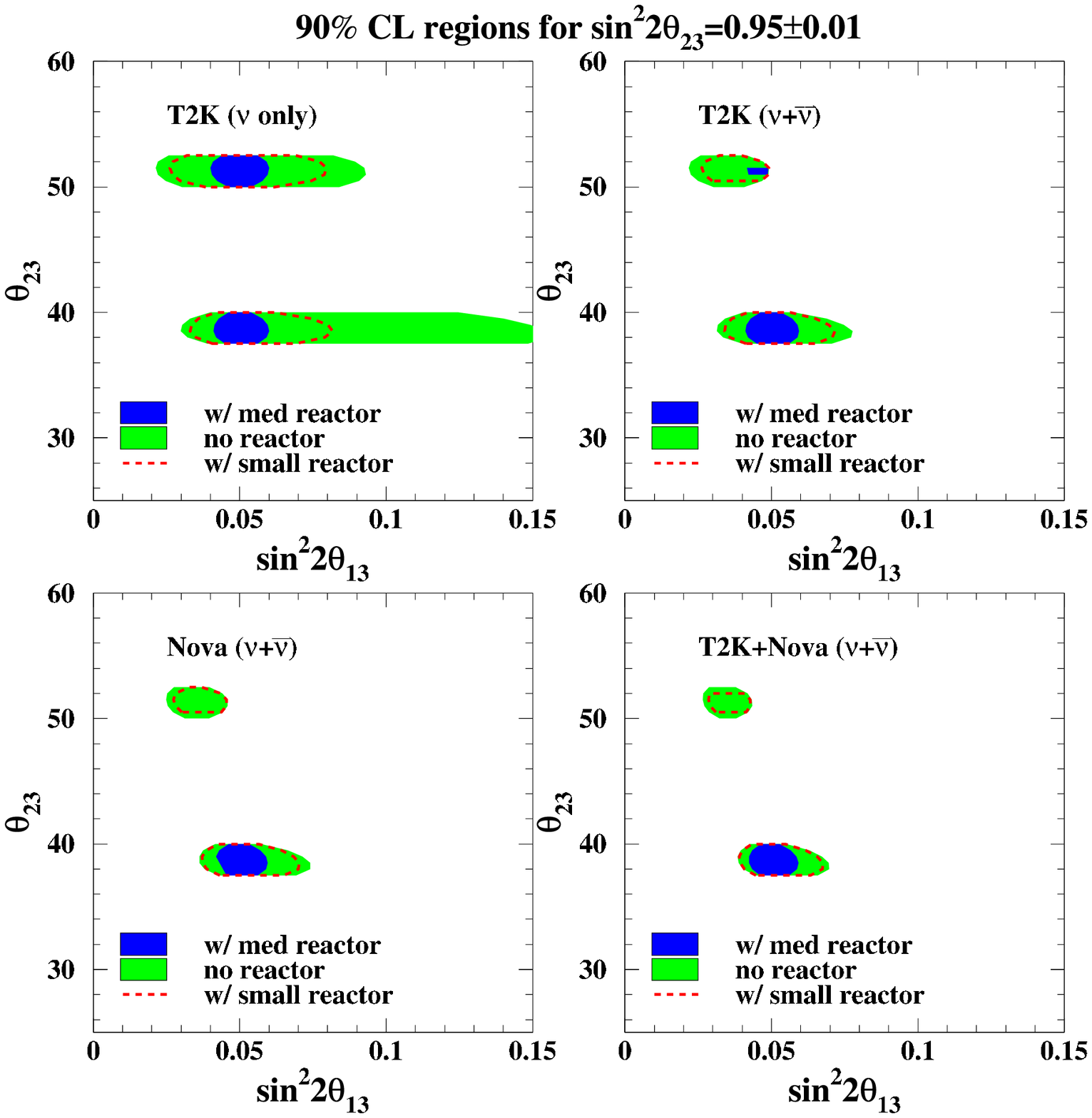}
\end{center}
\caption{90\% C.L. allowed regions for simulated data with an underlying
oscillation parameters of $\sin^{2}2\theta_{13}=0.05$, 
$\theta_{23}=38.54^{\circ}$,
$\Delta m^{2}=2.5\pm 0.1\times10^{-3}$ eV$^{2}$ and $\delta_{CP}=270^{\circ}$. 
The analysis
includes the restriction that $\sin^{2}2\theta_{23}=0.95\pm0.01$. The grey
regions are for various long-baseline combinations of the T2K and Nova
experiments for three year running periods. The black regions are the 90\%
C.L. allowed regions for a combined medium reactor plus long-baseline
analysis. The dashed lines indicate a combined analysis with a small
reactor measurement.}%
\label{23vs13_015}%
\end{figure}

Future measurements of $\nu_{\mu}$ disappearance in the atmospheric $\Delta
m^{2}$ region by the T2K or Nova experiments could reduce the uncertainty in
$\sin^{2}2\theta_{23}$ to order 0.01 \cite{t2kloi}, but would still leave the
$\theta_{23}$ vs. $\frac{\pi}{2}-\theta_{23}$ ambiguity. Figure
\ref{23vs13_015} shows the different combinations of long-baseline
results with or without the inclusion of a medium scale reactor measurement 
for $\sin^{2}2\theta_{23}=0.95\pm0.01$. Again, the medium scale reactor 
resolves the
degeneracy when combined with T2K and Nova neutrino and antineutrino
running, whereas the small reactor does not.

\begin{figure}[ptb]
\begin{center}
\includegraphics[
width=6.5in
]{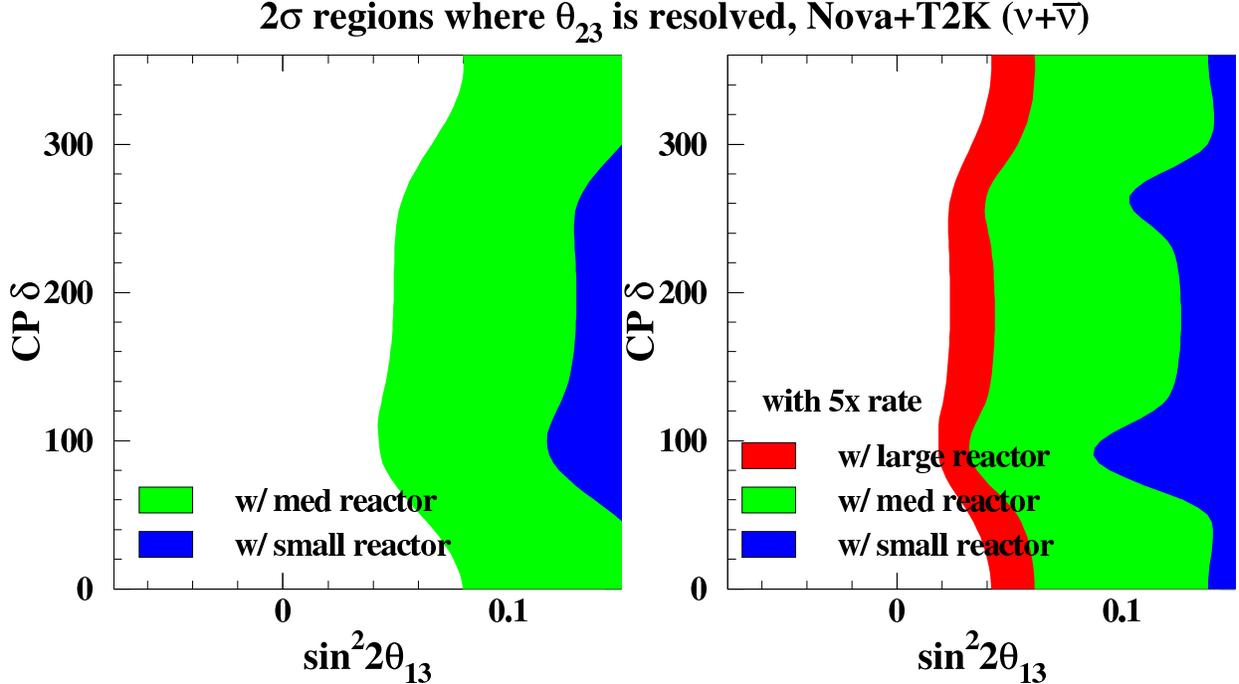}
\end{center}
\caption{Regions in the $\delta _{CP}-\sin ^{2}2\theta _{13}$ plane
  for which the ambiguity for $\theta _{23}>45^{\circ }$
  or$\,<45^{\circ }$ is resolved at
the two standard deviation level for $\sin ^{2}2\theta _{23}=0.95\pm
  0.01$.
(The allowed regions are to the right of the curves.)
The left plot is the result for a combined data set using the nominal rate
T2K and Nova data with medium (light grey) or small (black) scale
reactor data. The right plot shows similar results for the enhanced rate $%
(\times 5)$ T2K and Nova data combined with a small (black), medium
(light grey), and large (grey) scale reactor data.}%
\label{th23resolve}%
\end{figure}

In general, the regions of the $\delta $ versus $\sin ^{2}2\theta _{13}$
plane where this degeneracy can be resolved at the two standard deviation
level is shown in Fig. \ref{th23resolve}.
 From the figure, it is seen that resolving this ambiguity puts a premium on
including reactor data with good precision. Combined with the upgraded
off-axis data, a medium (large) scale reactor experiment can cover most of
the parameter space for $\sin ^{2}2\theta _{13}$ values greater than 0.05
(0.025). If $\sin ^{2}2\theta _{23}$ turns out too much different from 1.0,
resolving this ambiguity will become very important to making progress
toward determining the CP phase and the mass hierarchy.

\section{Constraining the CP Violation parameters at 90\% C.L. or discovery at
  2$\sigma$ or 3$\sigma$}

One of the important goals of an oscillation physics program is to determine
if CP violating effects are present in the lepton sector, as probed through the
neutrino mixing matrix. In contrast to the reactor disappearance probability,
the oscillation probabilities for the long-baseline experiments are affected
by the value of the CP violation phase $\delta_{CP}$. Due to these differences,
 combinations of long-baseline neutrino, antineutrino, and
reactor measurements can be used to isolate these CP violating effects and
place constraints on $\delta_{CP}$ (see Ref. \cite{Minakata2003}). The
size of these effects is scaled by
the value of $\sin^{2}2\theta_{13}$ which is therefore an important parameter
for setting the sensitivity to CP violation. The analysis also needs to
include the uncertainties associated with the other parameters, especially
the mass hierarchy. 

\begin{figure}[ptb]
\begin{center}
\includegraphics[
width=6.5in]
{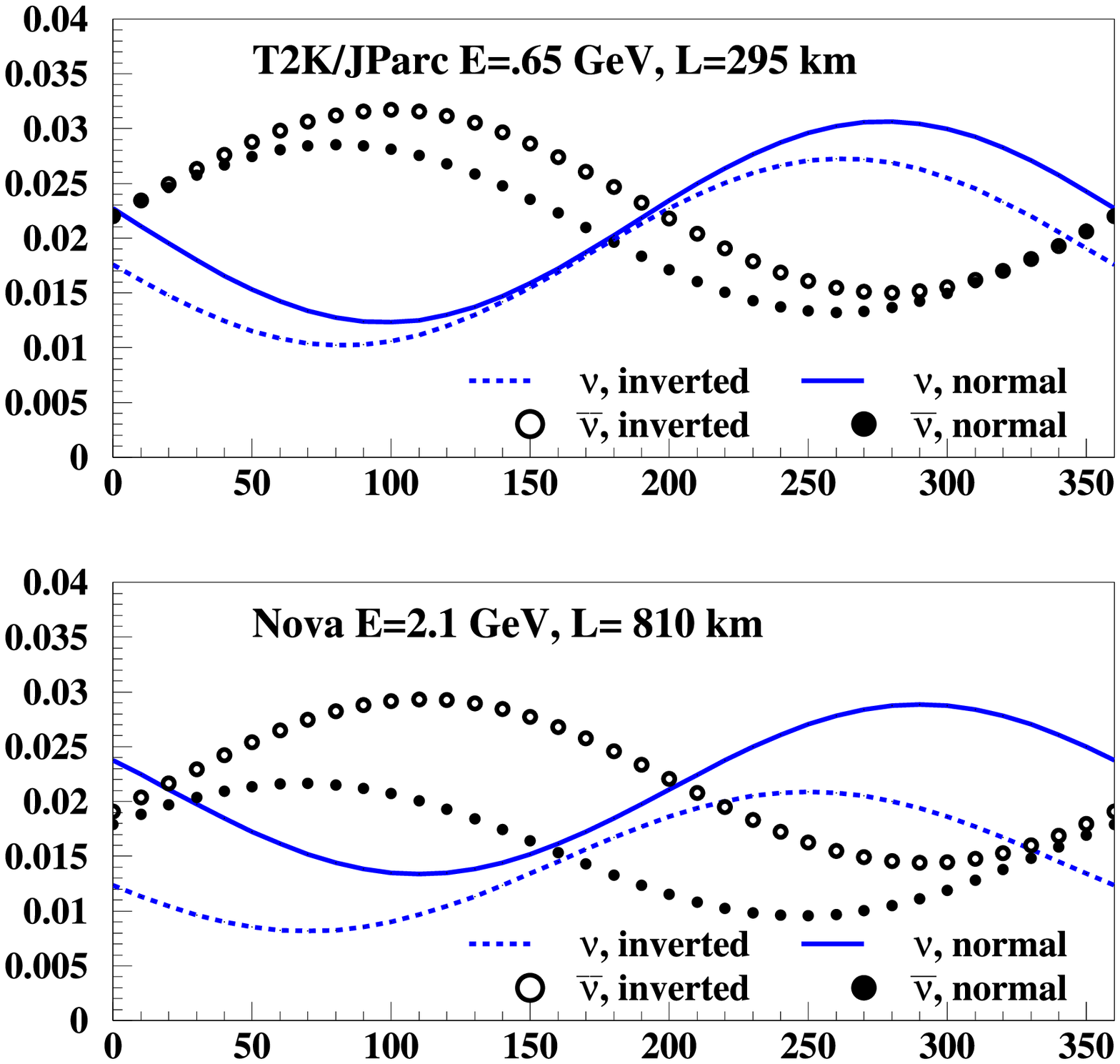}
\end{center}
\caption{Oscillation probability for $\nu_{\mu}\rightarrow\nu_{e}$ appearance
vs. $\delta_{CP}$ for the T2K (top) and Nova (bottom) experimental setups with
$\Delta m^{2}=2.5\times10^{-3}$ eV$^{2}$, $\sin^{2}2\theta_{13}=0.05$ and 
$\sin^{2}2\theta_{23}=0.95$. The four
curves correspond to pure neutrino (solid: normal hierarchy, dashed: inverted hierarchy) or antineutrino (black circle: normal hierarchy, white circle: inverted hierarchy) beams.}%
\label{lbl_oscprob}%
\end{figure}

Figure \ref{lbl_oscprob} gives the $\nu_{\mu}%
\rightarrow\nu_{e}$ appearance oscillation probability as a function of
$\delta_{CP}$ for the various combinations of beam type and mass hierarchy for
$\sin^{2}2\theta_{13}=0.05$.  As seen from the figure, there is a dramatic
difference between the T2K and Nova experiments due to matter effects. The goal
of a combined oscillation analysis would be to use these differences in the 
oscillation probabilities to constrain the CP violation phase and the 
mass hierarchy. For these analyses, reactor measurements provide an 
unambiguous constraint on the value of $\sin^{2}2\theta_{13}$.

\begin{figure}[ptb]
\begin{center}
\scalebox{1.}{\includegraphics[
width=6.5in
]{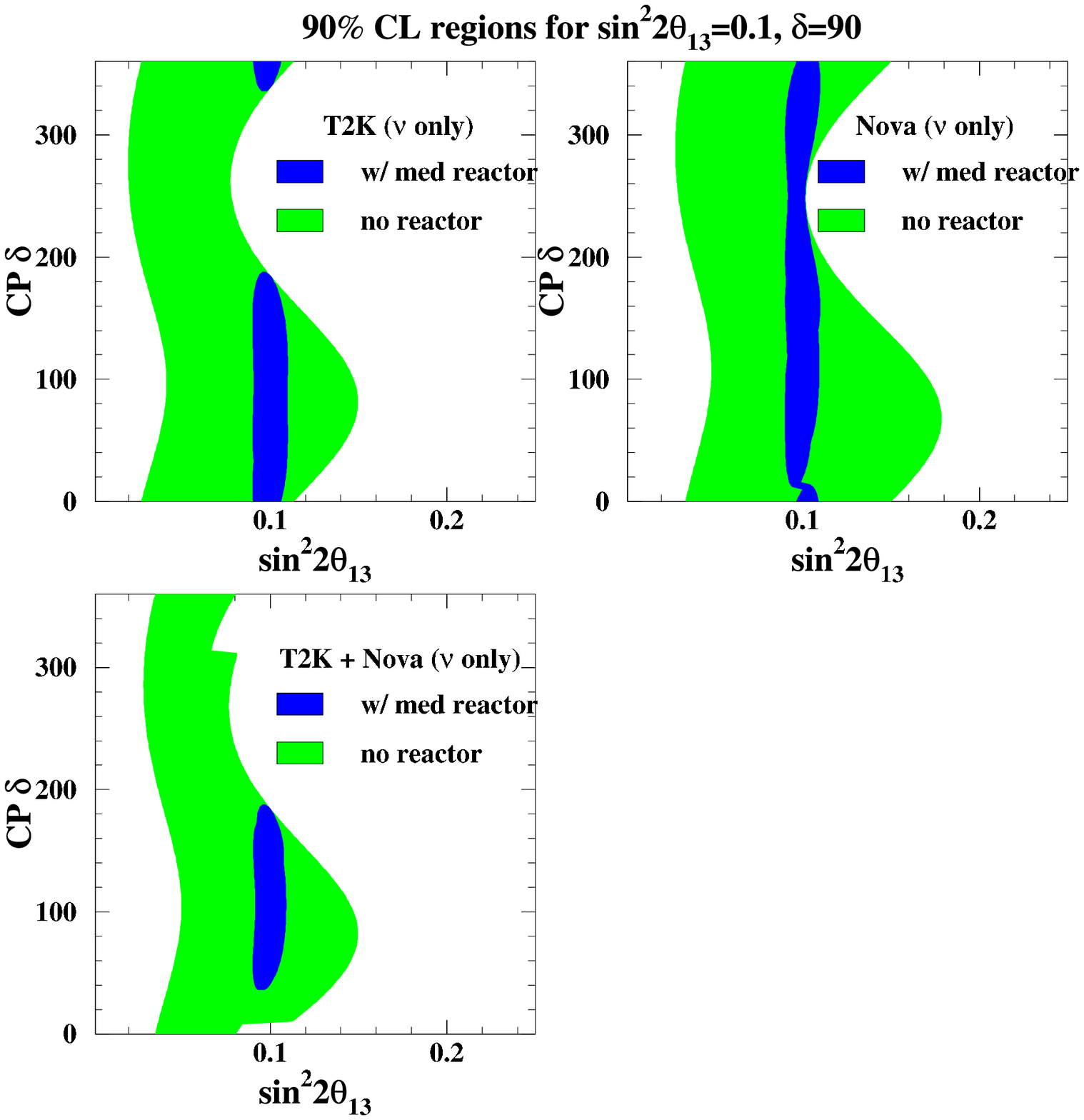}}
\end{center}
\caption{90\% C.L. regions for underlying oscillation parameters of $\sin
^{2}2\theta_{13}=0.1$, $\sin^{2}2\theta_{23}=0.95\pm0.01$, 
$\Delta m^{2}=2.5\pm 0.1\times10^{-3}$ eV$^{2}$ and $\delta_{CP}=90^{\circ}$.
 The grey regions are for the T2K, Nova or Nova+T2K experiments for 
five years of neutrino running. The black regions are 
the
90\% C.L. allowed regions for a combined medium reactor plus long-baseline
analysis. The shelf-like discontinuity around $\delta_{CP}=0^{\circ}$ and $360^{\circ}$ are due to the ability to distinguish the mass hierarchy at those values of $\delta$.  }%
\label{5yrdcp90}%
\end{figure}

From Figure \ref{lbl_oscprob}, it can also
be seen that a measurement of the appearance
probability for neutrino running alone could give information on $\delta_{CP}$
if the value of $\sin^{2}2\theta_{13}$ was known with sufficient accuracy 
from a reactor
oscillation measurement. For high values of $\sin^{2}2\theta_{13}$ near the
current limit, Figure \ref{5yrdcp90} shows how 
$\delta_{CP}$ can be constrained with
neutrino only running for Nova, T2K, and the combination
of the two with and without a medium reactor. For the true value of 
$\delta_{CP} = 90^{\circ}$,
T2K and the reactor narrows down the allowed range in $\delta_{CP}$.

\begin{figure}[ptb]
\begin{center}
\includegraphics[
width=6.5in
]{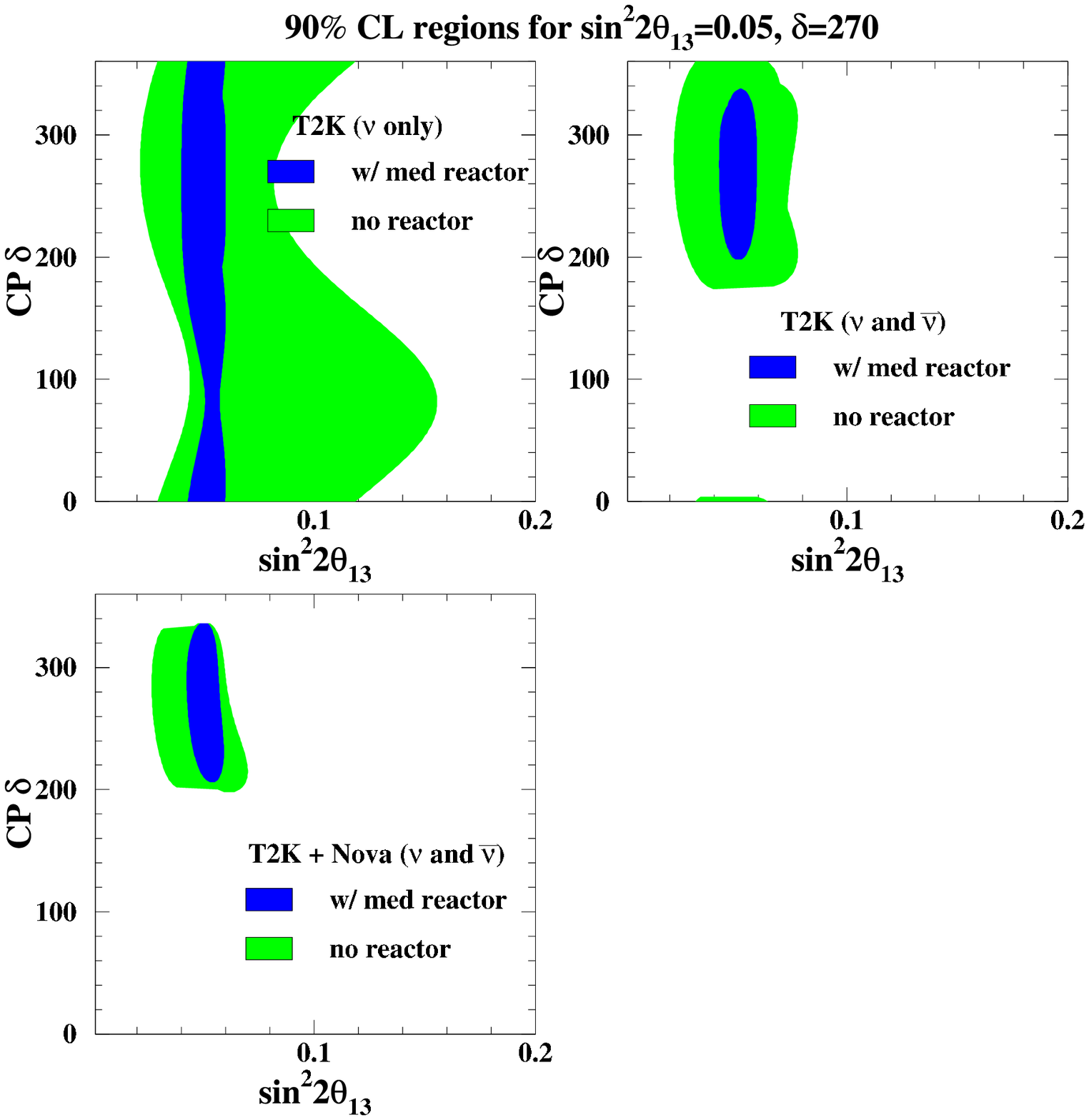}
\end{center}
\caption{90\% C.L. regions for various combinations of oscillation results for
$\sin^{2}2\theta_{13}=0.05$ and $\delta_{CP}=270^{\circ}$ 
($\Delta m^{2}=2.5\pm 0.1\times10^{-3}$ eV$^{2}$, and 
$\sin^{2}2\theta_{23}=0.95\pm0.01$). Upper Left: 3 year $\nu
$--only T2K data with (black) and without (grey) medium scale reactor results. 
Upper Right: T2K $\nu+\overline{\nu}$ with (black) and without (grey) reactor 
result. Lower Left: T2K $\nu +\overline{\nu}$ plus Nova $\nu+\overline{\nu}$ 
with (black) and without (grey) reactor result.}%
\label{cpvs13_270}%
\end{figure}

Figure \ref{cpvs13_270} shows how the combinations
of various measurements, including antineutrino measurements, 
can be used to constrain the allowed CP violation
phase. The results are for a scenario with $\sin^{2}2\theta_{13}=0.05$ and the
optimum phase point $\delta_{CP}=270^{\circ}$ (where the difference
between the neutrino and antineutrino oscillation probability is the
largest for the normal hierarchy). In the upper left plot, a T2K $\nu
$--only (3 years) measurement is displayed first without any reactor
measurement (grey region), then combined with a medium scale reactor
measurement (black region). The upper right plot then shows what
happens when both neutrinos (3 years) and antineutrinos (3 years) are used
with and without the reactor measurement. Finally, the lower left plot shows the
combination of T2K and Nova with and without the reactor result. The medium 
reactor measurement, in all cases, significantly reduces the
uncertainty on $\theta_{13}$.

\begin{figure}[ptb]
\begin{center}
\includegraphics[
width=6.5in]
{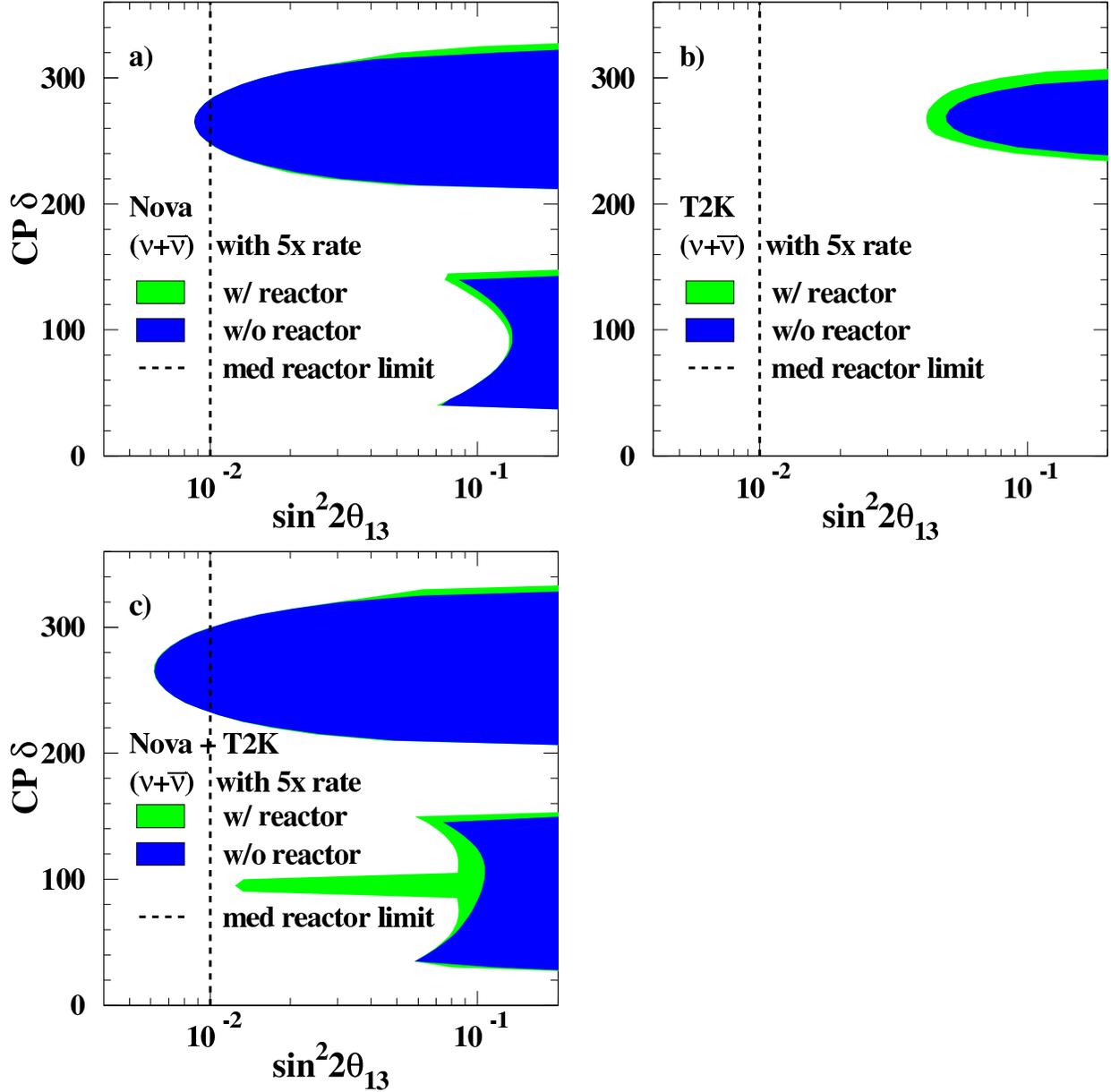}
\end{center}
\caption{Discovery regions in the 
$\delta_{CP}-\sin^{2}2\theta_{13}$ plane for which a
null CP ($\delta_{CP}$=0,$\pi$) violation solution is ruled out by at
least three standard deviations.
The black regions use long-baseline data only and the grey regions include data
from a medium scale reactor experiment. a) Nova ($\times$5 rate with
Proton Driver) $\nu$(3yr) + $\overline{\nu}$(3yr) data; b) T2K
($\times$5 rate) $\nu$(3yr) + $\overline{\nu}$(3yr)
; c) T2K ($\times$5 rate) $\nu$(3yr) +
$\overline{\nu}$(3yr) + Nova ($\times$5 rate) $\nu$(3yr) + $\overline{\nu}
$(3yr) data. (Nova,
T2K and Nova plus T2K with nominal rates is not shown since those
combinations are not capable of making a CP violation
discovery at the three standard deviation level.) The vertical dashed line indicates the 90\% CL upper limit for a medium reactor experiment alone.}%
\label{cp3sig}%
\end{figure}

As a measure of how well the CP phase can be constrained in general, Figure
\ref{cp3sig} gives the discovery 
regions in the $\delta_{CP}-\sin^{2}2\theta_{13}$ plane
for which a null CP violation solution ($\delta_{CP}=0$ or $\pi$) is ruled out 
by at least three standard
deviations. The black regions use long-baseline data only and the grey regions include data from a medium scale reactor experiment. 
Plot a) is for Nova only, b) is for T2K only and c) includes data from both
T2K and Nova. 
(The results for nominal beam rates are not shown since they 
do not provide any restrictions at the three standard deviation level.)
From the plots, it is seen that the combination of T2K and Nova with increased 
intensity can start to probe
the CP violation phase space if $\sin^{2}2\theta_{13}\gtrsim0.02$. (The narrow
region in the lower range of $\delta$ is due to the ambiguity between normal 
and inverted hierarchies.)
The reactor measurements show how viable a CP violation
measurement will be with the various combinations of the long-baseline setups.

\section{Determining the Mass Hierarchy}

For constraining the mass hierarchy, one needs to compare measurements in a
region where the oscillation probability changes significantly for a normal
versus inverted mass spectrum (see Figure \ref{lbl_oscprob}). 
These type of changes can be induced by matter effects as the neutrinos or
antineutrinos propagate through material.
The Nova experiment is particularly important here due to the 
long distance the neutrinos travel through the matter of the earth. 

\begin{figure}[ptb]
\begin{center}
\includegraphics[
width=6.5in]
{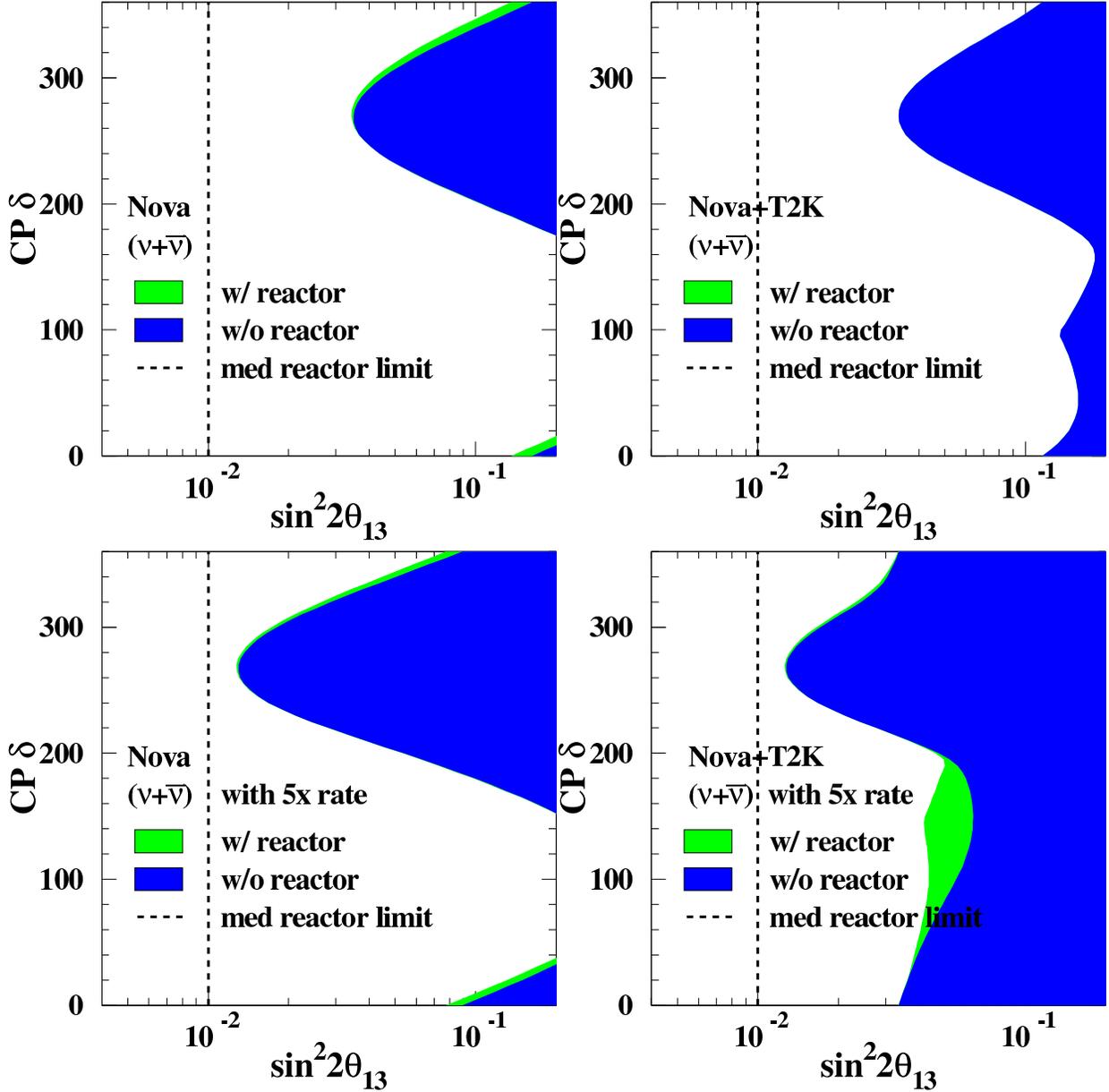}
\end{center}
\caption{Regions in the $\delta_{CP}-\sin^{2}2\theta_{13}$ plane for which the
mass hierarchy is resolved by two standard deviations. The grey regions use
long-baseline data only and the black regions add data from a medium scale
reactor experiment. a) Nova $\nu$(3yr) +
$\overline{\nu}$(3yr) data; b) Nova plus T2K with $\nu$(3yr) + $\overline{\nu
}$(3yr) data; c) Nova ($\times$5 beam rate) with $\nu$(3yr) + $\overline{\nu}%
$(3yr) data; d) T2K ($\times$5 beam rate) $\nu$(3yr) + $\overline{\nu}%
$(3yr) + Nova ($\times$5 beam rate) $\nu$(3yr) + $\overline{\nu}$(3yr)
data. 
The vertical dashed line indicates the 90\% CL upper limit for a medium reactor experiment alone.
}%
\label{hier2sig}%
\end{figure}

An accurate determination of the hierarchy is possible by combining the
results from long-baseline neutrino and antineutrino data. Here again, the
ambiguity with respect to the value of $\delta_{CP}$ limits the determination
to regions in the $\sin^{2}2\theta_{13}-\delta_{CP}$ plane. Figure
\ref{hier2sig} shows the regions for which the mass hierarchy is resolved by
two standard deviations. The black regions use long-baseline data only and the
grey regions add data from a medium scale reactor experiment.

These plots show that the mass hierarchy can be determined for limited regions
with $\sin^{2}2\theta_{13}>0.05$ at the nominal beam rates.  With the enhanced
($\times 5$) rates, the combination of T2K plus Nova covers most of the 
$\delta_{CP}$ range for $\sin^{2}2\theta_{13}>0.03$. Note that in the upper right plot the black and grey regions lie on top of each other, that is, a reactor doesn't contribute for a short joint run of Nova and T2K but does for a longer one.

\section{Other Studies}

\begin{figure}[ptb]
\begin{center}
\includegraphics[
width=6.5in]
{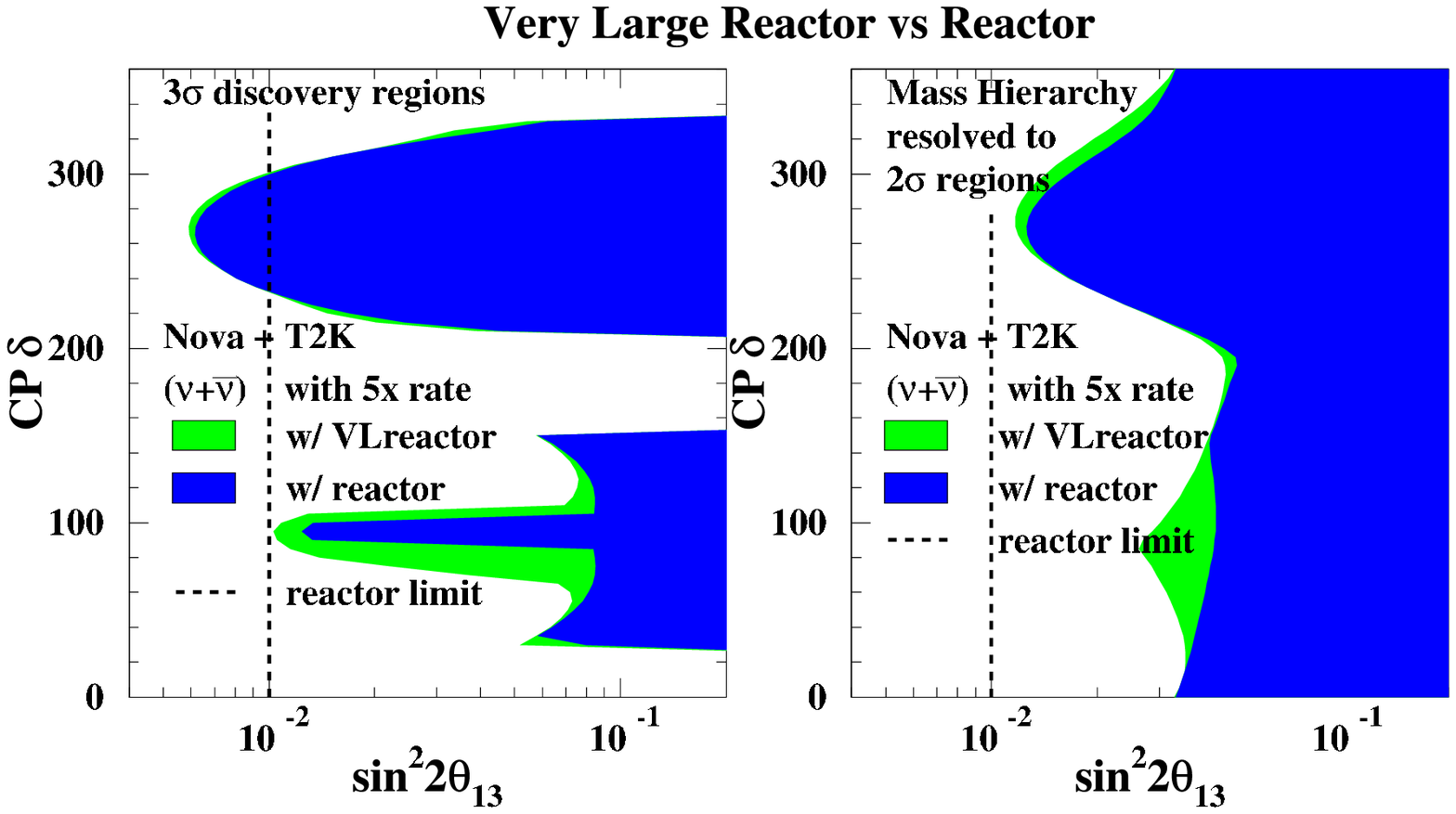}
\end{center}
\caption{Regions in the $\delta_{CP}-\sin^{2}2\theta_{13}$ plane where null 
CP violation is ruled out by three standard deviations (left plot) and where the
mass hierarchy is resolved by two standard deviations (right plot) for
a very large (VL) reactor, with sensitivity of $\sin^2 2\theta_{13}> 0.003$ at
90\% C.L.  Both plots are for
 Nova and T2K ($\times$5 beam rate) with $\nu$(3yr) + $\overline{\nu}%
$(3yr) data. The black region indicates a measurement including the medium sized reactor, and the grey region includes the very large reactor. 
 The vertical dashed line indicates the 90\% CL upper limit for a medium reactor experiment alone.
}
\label{ultre}%
\end{figure}

It may be possible in the future to use a very large detector at a site 
with multiple reactors
to push the $\sin^{2}2\theta_{13}$ sensitivity down even beyond the 
``large reactor'' scenario to the level of 0.003 at 90\% C.L.  
Combining such a reactor
measurement with enhanced long-baseline results improves coverage minimally for a CP measurement and hierarchy determination as compared to a medium
reactor, as seen in Figure \ref{ultre}.

\begin{figure}[ptb]
\begin{center}
\scalebox{1.}{\includegraphics[
width=6.5in]
{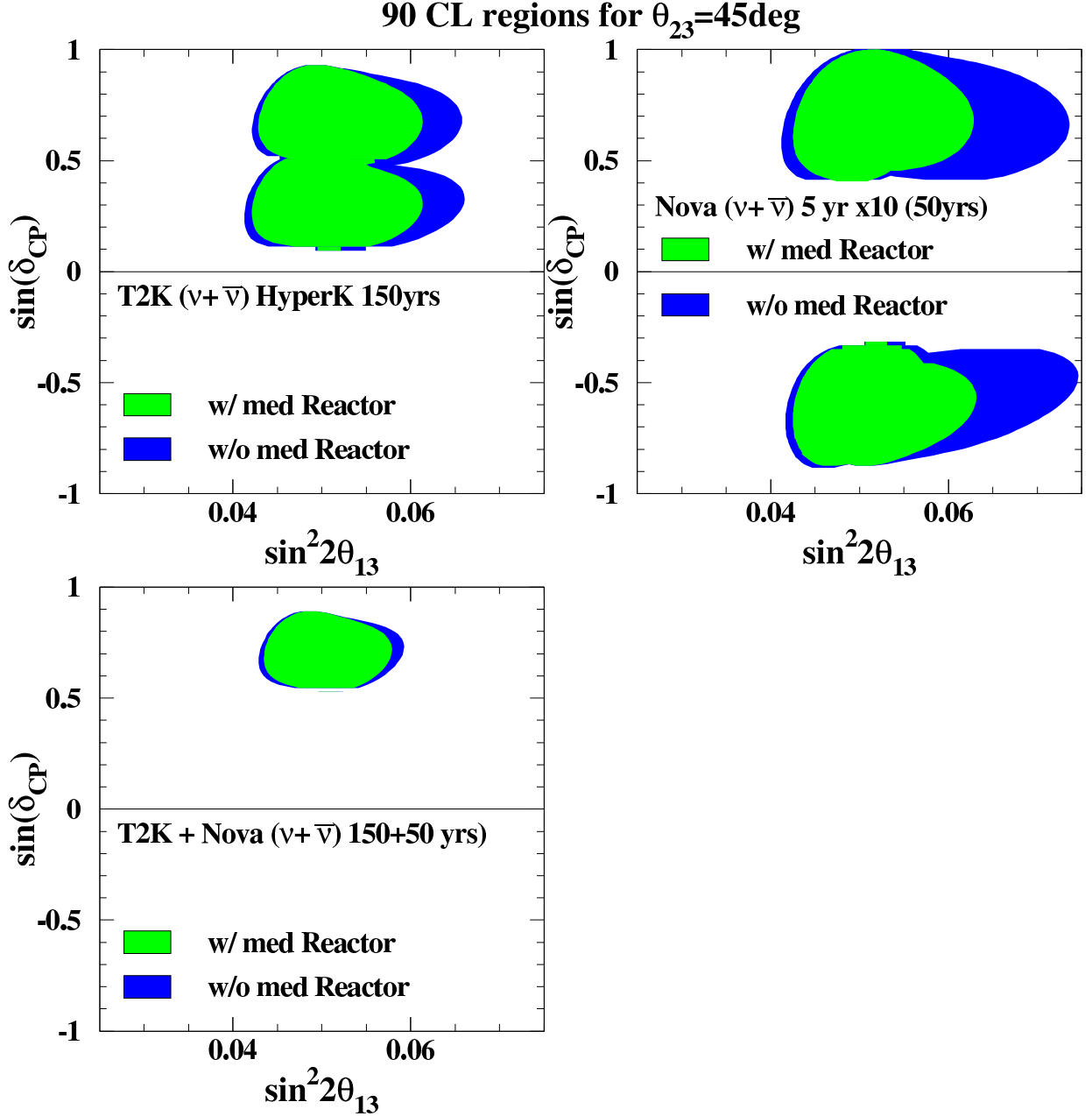}}
\end{center}
\caption{90\% C.L. allowed regions in the $\sin\delta_{CP}-\sin^{2}2\theta_{13}$
 plane for T2K (upper left) and Nova (upper right) and 
T2K (150yrs) and Nova (50yrs) (lower left) shown with the reactor measurement (in grey) and without (in black) all for equal $\nu$ and $\overline{\nu}$ running. $\theta_{23}$ and $\Delta m^{2}$ are allowed to vary 
within future uncertainties. $\theta_{23}=45^{\circ}$. T2K and Nova combined (lower left plot) correctly resolves the mass hierarchy.}%
\label{parke1}%
\end{figure}

\begin{figure}[ptb]
\begin{center}
\scalebox{1.}{\includegraphics[
width=6.5in]
{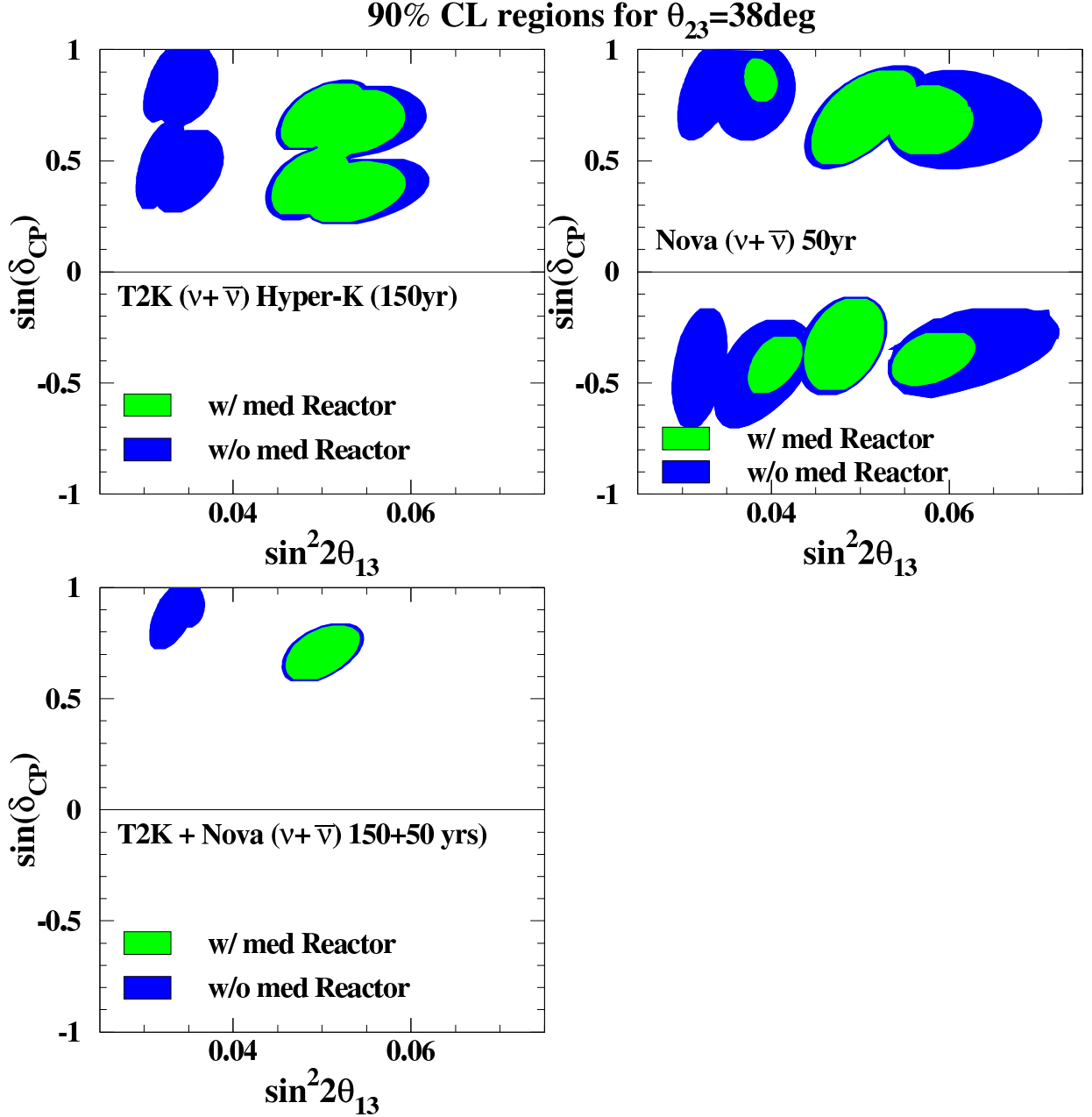}}
\end{center}
\caption{90\% CL allowed regions in the $\sin \delta_{CP} - \sin^{2}2\theta_{13}$ plane for T2K (upper left) and Nova (upper right) and 
T2K (150yrs) and Nova (50yrs) (lower left) shown with the reactor measurement (in grey) and without (in black) all for equal $\nu$ and $\overline{\nu}$ running. $\theta_{23}$ and $\Delta m^{2}$ are allowed to 
vary within future uncertainties. $\theta_{23}=38.54^{\circ}$. T2K and Nova combined
 (lower left plot) correctly resolves the the mass hierarchy only with the additon 
of a reactor experiment.}%
\label{parke2}%
\end{figure}

Figures  \ref{parke1} and \ref{parke2} show $\sin\delta_{CP}$ vs $\sin^{2}2\theta_{13}$ for future T2K and Nova upgrades. For this study, we consider a Hyper-K upgrade data sample to be equivalent to 150 years ($\times 20$ fiducial mass for 15 years) of T2K running and a Nova Phase II with a proton driver to be equivalent to 50 years ($\times 5$ increased beam rate or detector volume for 10 years) of normal Nova running. These plots, inspired by an earlier paper of Mena and Parke's \cite{Mena:2004sa}, show clearly the regions of degeneracy.

Ambiguities associated with the mass hierarchy produce multiple solutions and ambiguities associated with $\theta_{23}$ give extended regions along the $\theta_{13}$ direction. If $\theta_{23} =$ 45 degrees, the combination of T2K and Nova correctly determines the mass hierarchy, as shown in Figure \ref{parke1}, and gives only one allowed parameter region. On the other hand, with $\theta_{23}$ non maximal ($\theta_{23}=38.54^{\circ}$), Figure \ref{parke2} shows two allowed regions due to the $\theta_{23}$ ambiguity that can only be resolved with the addition of a reactor measurement.

\section{Conclusions}

Over the next decade, many experiments are planned to address neutrino 
oscillations and make improved measurements of the relevant parameters
of neutrino mixing.
As in the past, combining the results from different types of experiments 
with different setups will be necessary to map out the underlying
physics.  In the studies described here, we have tried to investigate the
sensitivity using the suite of currently planned or proposed experiments. 
The complementarity of a program with several long-baseline accelerator
experiments and several reactor measurements is clear and, if parameters are
favorable, will lead to significant progress in the understanding of neutrino
mixing and masses.

As part of this program, reactor measurements hold the promise 
of constraining or measuring the
$\theta_{13}$ mixing parameter and helping to resolve the ambiguity in
determining the
$\theta_{23}$ mixing parameter. The sizes of these parameters are important
inputs for models of lepton mass and mixing that span the range from GUTs
trying to relate the CKM and MNS matrix to extra dimension models that have
neutrinos propagating in the bulk. The smallness of $\theta_{13}$ relative to
the other angles may give a hint as to what the underlying theory may
be. Besides leading to a better understanding of neutrino mixing, these
angles, $\theta_{13}$ and $\theta_{23}$, are two of the twenty-six parameters
of the standard model and, as such, are worthy of high precision measurement
independently of other considerations. For $\theta_{13}$, a two detector
reactor experiment unambiguously measures the size of this angle with
significantly better precision than any other proposed experimental technique.
In addition, reactor data may be key for resolving the $\theta_{23}$ degeneracy.

Looking towards probing CP violation and the mass hierarchy in the 
neutrino sector, the field will need several high-rate, long-baseline experiments. 
Here, the size of $\theta_{13}$ will be important for interpreting the results
and for planning a viable future neutrino oscillation
program. In addition to setting the scale for future studies, a reactor result
when combined with long-baseline measurements may also give early constraints
on CP violation and early indications of the mass hierarchy. 
In the longer term, a combination of long-baseline experiments such as
T2K and Nova will start to give some 
information about these effects if $\sin^{2}2\theta_{13}>0.05$ and, with some upgrades and enhanced beam rates, will give good coverage if $\sin^{2}2\theta_{13}>0.03$.
If $\theta_{13}$ turns out to be smaller
than these values, one will need other strategies for getting to the physics.
Thus, an unambiguous determination $\theta_{13}$ from,
for example, a medium scale reactor experiment, is an important
ingredient in planning the strategy for this program, as well as accessing the
phenomenology of neutrino mixing.

%\begin{acknowledgments}
%need acknowlegdments
%\end{acknowledgments}

%\newpage 
\bibliography{reactor_offaxis}% Produces the bibliography via BibTeX.

\end{document}